\documentclass[sn-mathphys,Numbered]{sn-jnl}

\usepackage{graphicx}%
\usepackage{multirow}%
\usepackage{amsmath,amssymb,amsfonts}%
\usepackage{amsthm}%
\usepackage{mathrsfs}%
\usepackage[title]{appendix}%
\usepackage{xcolor}%
\usepackage{textcomp}%
\usepackage{manyfoot}%
\usepackage{booktabs}%
\usepackage{algorithm}%
\usepackage{algorithmicx}%
\usepackage{algpseudocode}%
\usepackage{listings}%

\usepackage{bm}

\title{2D Signal Estimation for Sparse Distributed Target Photon Counting Data}

\author*[1]{\fnm{Matthew} \sur{Hayman}}\email{mhayman@ucar.edu}

\author[1]{\fnm{Robert A.} \sur{Stillwell}}\email{stillwel@ucar.edu}

\author[1]{\fnm{Josh} \sur{Carnes}}\email{jcarnes@ucar.edu}

\author[2]{\fnm{Grant J.} \sur{Kirchhoff}}\email{grant.kirchhoff@colorado.edu}

\author[1]{\fnm{Scott M.} \sur{Spuler}}\email{spuler@ucar.edu}

\author[2]{\fnm{Jeffrey P.} \sur{Thayer}}\email{Jeffrey.Thayer@colorado.edu}

\affil*[1]{\orgdiv{Earth Observing Lab}, \orgname{National Center for Atmospheric Research}, \orgaddress{\street{PO Box 3000}, \city{Boulder}, \postcode{80307}, \state{CO}, \country{USA}}}

\affil[2]{\orgdiv{Ann and H.J. Smead Aerospace Engineering Sciences}, \orgname{University of Colorado at Boulder}, \orgaddress{\street{429 UCB, 3775 Discovery Drive}, \city{City}, \postcode{80303}, \state{CO}, \country{USA}}}

\abstract{
In this study, we explore the utilization of maximum likelihood estimation for the analysis of sparse photon counting data obtained from distributed target lidar systems. Specifically, we adapt the Poisson Total Variation processing technique to cater to this application. By assuming a Poisson noise model for the photon count observations, our approach yields denoised estimates of backscatter photon flux and related parameters. This facilitates the processing of raw photon counting signals with exceptionally high temporal and range resolutions (demonstrated here to 50 Hz and 75 cm resolutions), including data acquired through time-correlated single photon counting, without significant sacrifice of resolution. Through examination involving both simulated and real-world 2D atmospheric data, our method consistently demonstrates superior accuracy in signal recovery compared to the conventional histogram-based approach commonly employed in distributed target lidar applications.}

\begin{document}

\flushbottom
\maketitle
\thispagestyle{empty}

\section*{Introduction}

Photon counting is commonly employed in distributed target lidar where photon fluxes are low.  This arises from various factors, including the range of the volume under interrogation, inefficient scattering or collection mechanisms, and limitations imposed by instrument parameters (e.g. low pulse energy, small collection aperture resulting from eye-safety, size, weight, and power restrictions and low efficiency detection processes).  In these applications, a crucial step is estimating the amplitude of the backscatter optical signal -- a key intermediary of any derived lidar variable where errors in the backscatter estimation problem can only compound and propagate to the derived variables.   Traditionally, backscatter estimation is performed by placing observed photon counts into histograms of fixed range and time resolution.  The selection of bin widths, and consequently time and range resolutions, are determined based on resolvable signal levels.  Larger bin width allows for greater noise suppression but comes at the cost of reduced resolution -- a trade-off that must be carefully considered.

The standard approach to processing lidar data treats these histogrammed photon counts as the true backscatter signal.  This assumption is typically valid under two conditions: (1.) backscatter photon counts in a bin are high, resulting in minimal shot noise, and (2.) when the instantaneous photon arrival rate is small throughout the bin interval (negligible nonlinear detector effects are present).  In this work, we focus on addressing cases where the first condition (1.) is invalid due to significant shot noise.  This situation occurs when we aim to observe structure at high resolution,  a relative term indicated by the relationship between the capture time of a bin and its associated mean photon flux.  Consequently, the absolute photon counts in any given bin becomes relatively small.  While the second condition (2.) is also important, some issues related to nonlinear effects have been addressed for hard target lidar in previous works\cite{Rapp2019,Rapp2021}.   We leave it to future investigations to adapt these approaches for distributed target lidar.

In placing photon counts in a histogram, the expected number of observed photons $\alpha_i$ in a bin $i$ is given by
\begin{equation}
\alpha_i = \sum_{n=1}^N \int_{t_i}^{t_i+\Delta t}\rho_n(\tau) d \tau
\end{equation}
where $\rho_n(t)$ is the photon flux at time-of-flight $t$ for laser shot $n$, the bin number is defined by the interval $[t_i, t_i+\Delta t)$ and $\Delta t$ is the histogram bin width in time-of-flight.  A total of $N$ laser shots are acquired per histogram.  In this work we let $t$ be time-of-flight relative to the most recent emitted laser pulse.  

The standard approach to estimating backscatter amplitude is to place photon arrivals in a histogram (either immediately or during post-processing) and the observed photon counts in a bin is a Poisson random variable
\begin{equation}
    y_{i} \sim Poisson(\alpha_i).
\end{equation}
Given these photon count observations, the ubiquitous standard estimator for the backscatter amplitude is
\begin{equation}
    \tilde{\alpha}_i \approx y_i
\end{equation}
or if the photon flux is desired
\begin{equation}
    \tilde{\rho}_i \approx y_i/(N\Delta t).
\end{equation}

Because the photon counts in the histogram bin are Poisson distributed, the standard deviation in the observed counts is $\sigma_{y_i} = \sqrt{\alpha_i}$.  If the signal-to-noise (SNR) is defined as the signal mean divided by standard deviation, then for a given bin
\begin{equation}
    SNR_i = \sqrt{\alpha_i} = \sqrt{N\Delta t \bar{\rho}_i}
\end{equation}
where $\bar{\rho}_{i}$ is the mean photon flux over the acquisition interval of the histogram defined
\begin{equation}
\bar{\rho}_{i} = \frac{1}{N\Delta t} \sum_{n=1}^N \int_{t_i}^{t_i+\Delta t}\rho_n(\tau) d \tau = \frac{\alpha_i}{N\Delta t}.
\end{equation}

Operating under the assumption that the photon flux remains constant in the region of observation (although this assumption may not strictly hold for many operational remote sensing observations), it becomes evident that the precision of the standard estimator is directly linked to the histogram bin size.  This connection demonstrates the problem of applying the standard estimator to high-resolution photon count data, where the reduction in SNR resulting from the use of small histogram bins and short integration times can only be mitigated by increasing the photon flux.  Conventionally achieving higher photon flux entails employing higher laser power and larger collection apertures.  However, this will tend to increase instrument cost, size, weight, and power (SWaP) requirements, making systems more complex.  Additionally, it can push photon fluxes into nonlinear detection regimes, potentially compromising data accuracy and reliability.

It is important to recognize that as we push for higher resolutions, the precision of the standard estimator diminishes.  However, it is equally important to recognize that the high-resolution data has no less information than that at coarser resolutions; in fact, the converse is true.  When we bin data into coarser resolution information is lost.  Moreover, the information content of high-resolution data is increasingly contained in the spatial clustering of those counts instead of the histogram counts. This limitation of the standard estimation approach becomes especially critical when dealing with sparse data, where the ability to leverage clustering information is essential for accurately estimating the backscatter signal.

For the purposes of this work, a sparse scene is characterized by instances in which $\alpha_i < 1$, or equivalently $\bar{\rho}_i  < 1/(N \Delta t)$, such that the probability of zero photon counts is relatively high and the probability of more than one photon count is relatively low.  Thus sparse data can be the product of very high-resolution data acquisition or very low photon flux, but the distinction is not critical for the applications of this work.

Time-tagged data acquisition, sometimes referred to as Time Correlated Single Photon Counting (TCSPC) is frequently employed in hard target lidar applications \cite{McCarthy2013,Pawlikowska2017}.  In this mode of acquisition, photon counts from a detector are stored as time stamps, recording time-of-flight relative to the time of the transmitted laser pulse.  This information is collected on a shot-by-shot basis.  The acquisition range resolution is determined by the precision of the onboard clock measuring the time-of-flight,  enabling the captured range resolutions to be extremely high (down to centimeters or millimeters).  In principle, when range capture resolution is infinitely fine, this approach avoids any information loss during the acquisition step.  This ideal scenario represents an extreme case where \emph{all} information content is carried in the clustering of the observed photon counts, with no information carried in the count quantity itself.  While this time-tagged acquisition avoids any loss of information, we should acknowledge that it also introduces practical challenges when implemented in regular operations.  One of the main challenges is the substantial data volumes generated by high-resolution data capture.  This often requires large data storage to archive data and high data transfer rates to move the raw data off the lidar device or platform.  This becomes particularly significant for instruments that run continuously for long time periods.

Seminal work in using time tag data for atmospheric lidar, a common application of distributed target lidar, was demonstrated in \cite{Barton-Grimley2018} and more recently \cite{Yang2023}.  These instruments have captured atmospheric backscatter data at unprecedented resolutions.  Consequently, they have the potential to provide important insights into the highly dynamic and heterogeneous nature of many atmospheric targets.  The influence of heterogeneous and dynamic targets on lidar data product accuracy have received relatively little attention, but there is some growing evidence that these factors can adversely affect remote sensing data.  For instance, one study has shown that averaging over heterogeneous cloud structures can bias CALIOP backscatter and depolarization data products\cite{Alkasem2017}.  Additionally, heterogenous cloud structure has been found to produce errors in estimates of cloud microphysical properties from passive satellite-based retrievals\cite{Arola2022}.  The ability to acquire high-resolution lidar data is an important step toward gaining a better understanding of these often ignored errors in real atmospheric scenarios.

Exploring high-resolution interrogation of the atmosphere is an important and largely unexplored scientific domain.  However, it is important to note that the current products of the aforementioned instruments still represent an incomplete solution to investigate this domain.  Although these works demonstrate high resolution \emph{capture} of photon count data, a key distinction lies between the resolution at which data is captured and the resolutions actually supported by the information content.  In essence, these instruments may simply be oversampling a noisy signal.  While the time-tagging approach seems to offer unprecedented resolution in the capture of atmospheric structure, the remaining challenge lies in actually recovering lidar signal estimates while still leveraging the benefits of the acquisition mode.  In both works mentioned, the high-resolution time-tagged data is eventually placed in histograms, and the standard estimation approach is employed to suppress shot noise and recover an estimated signal with acceptable precision.  However, this requires significant reductions in resolution, and therefore largely negates the benefit of the high-resolution acquisition approach employed by these instruments.  

Quantum Parametric Mode Sorting (QPMS) is a lidar technique often employing time tagging.  It leverages mode matching criteria in nonlinear photonic crystals to improve the SNR in the optical signals reaching the detectors, as outlined in \cite{Shahverdi2018,Zhu2021}.  While this technique may enable much better noise rejection, it comes at the cost of significantly lower total photon flux on the detectors due to the tightly constrained mode selection criteria.  As a result, lidar profiles generated using QPMS in time-varying scenes will almost inevitably be sparse.  It's worth noting that published works on this subject tend to refrain from providing details on the integration times compared to standard approaches and appear to largely focus on very close-range hard-target characterized by high backscatter.  This forces us to speculate regarding the possible fluxes expected in distributed target lidar.  While most experiments with QPMS have focused on laboratory demonstrations with static targets, the possible and proposed applications for the technology have included environmental sensing problems from moving remote platforms where distributed target lidar would necessarily be employed (e.g. see \cite{LeeAGU2022}).

When developing sensors and data processing techniques for environmental sensing it's crucial to acknowledge that the scenes under interrogation are rarely static.  This challenge becomes further compounded when the sensor is on a moving platform.  However, it is important to note that laboratory demonstrations are able to interrogate static scenes over time scales that do not accurately reflect many of the proposed real-world applications.  And in the instance of \cite{Yang2023} high \emph{range} resolution atmospheric backscatter estimates were achieved, but at the apparent expense of time resolution.  The highly dynamic nature of the clouds under investigation means that integration in time will smear out the image in range and largely negate the benefit of the sub-meter range resolution in the acquisition system.  To remain fundamentally true to the notion of high-resolution distributed target lidar, it is essential to perform an accurate retrieval from sparse photon counting data that accounts for variations in both time and range.  In order to maximize signal resolution, more advanced signal processing techniques are needed to recover useful signals from sparse photon counting data \emph{and} non-stationary scenes without dramatically sacrificing the signal resolution in either axis.  This is where Maximum Likelihood Estimation (MLE) can be employed to improve lidar data products by attempting to more efficiently leverage the information content, constraints, and uncertainties in the lidar signals.

The benefit of applying MLE is that it potentially enables signal recovery by leveraging both the photon counts and their clustering to generate a signal estimate.  MLE can thus be applied to time-tagged data directly, or similarly, to sparse and noisy histograms.  This approach represents a realistic means of achieving better observations, rather than degraded ones, as we capture, and subsequently process, data at higher resolutions.  While the raw observations appear noisy at high resolutions, the recovered signal estimates can still have low noise contamination, and more accurately reproduce the target structure compared to lower-resolution histogram estimator methods, which prioritize noise reduction at the expense of capturing accurate structure details.

Prior works on sparse photon counting problems have focused on single point target (hard target) ranging \cite{Altmann2016,Halimi2017}.  However, this does not directly translate to challenges posed by volumetric (distributed) target scenarios, where the backscatter amplitude serves as the parameter, or intermediate parameter, of interest and is nonzero throughout the range profile.  Such scenarios include the sensing of the atmosphere, snow characterization, and more.  The distinctive contribution of this work lies in addressing the specific challenges posed in distributed target lidar systems and therefore fulfills the signal processing needs of the environmental sensing techniques described above. 

Poisson Total Variation (PTV) is a regularized maximum likelihood estimator for processing distributed target lidar data.  The technique has been demonstrated in prior works \cite{Marais2016,Marais2022} where the focus is on retrieving data products at resolutions typical to atmospheric lidar, and photon counts are not sparse.  In those studies, there was no concerted effort to evaluate the impact of processing raw data at higher resolutions than were already typical of the standard processing approach.  Sparse data scenarios were intentionally avoided.

In this research endeavor, we show how a modified version of PTV can effectively process sparse data characterized by high temporal and range resolution. In this case we obtain backscatter estimates from the MicroPulse DIAL at resolutions as fine as 0.02 s x 0.75 m.  This raw lidar data includes but is not limited to time-tagged data.  Through simulations and analysis of actual atmospheric lidar data, we demonstrate that PTV is able to provide more accurate estimates of photon arrival rate at a higher resolution than the standard histogram approach.  This demonstrates its utility for processing time-tagging and other high-resolution sparse photon counting data.  By achieving this, PTV can serve as a useful tool for studying the atmosphere's heterogeneous and dynamic structure and its resulting impact on lidar data product accuracy.  Importantly, our work completes the necessary components to leverage hardware described in \cite{Barton-Grimley2018,Yang2023}.  Our results indicate that sparse retrievals result in more accurate backscatter estimates than those achieved using PTV at lower resolutions, suggesting that nearly all atmospheric lidar data is over-averaged.  However, the full impact of this over-averaging is still not known, highlighting the need for further investigations using this signal processing tool.

\section*{Poisson Time Tag Noise Model}
Photon arrival times are described by a Poisson point process parameterized by the photon arrival rate $\rho(s)$ where $s$ is the continuous time dimension (without reference to a laser shot event).  As such, a sequence of photon detection times $\lbrace s_i \rbrace_{i=1}^I$ (assuming an ideal detector), where a total of $I$ photon counts are detected, will be described by the probability distribution function (PDF) for a heterogeneous Poisson point process \cite{Snyder} 
\begin{equation}\label{Pois1D}
P\left(\lbrace S_i = s_i\rbrace_{i=1}^I\right) = \exp\left[-\int_{0}^{s_I}\rho(\tau) d\tau\right] \prod_{i=1}^{I} \rho(s_i).
\end{equation}
In the equation above, we describe the detection process as a 1D time series, where in a lidar, there may be periodic laser firing.  As such, there may be a desire to leverage correlation in the target structure across multiple laser shots.  We can accordingly rewrite equation \eqref{Pois1D}
\begin{equation}\label{Pois1Dprime}
P\left(\lbrace S_i = s_i \rbrace_{i=1}^{I} \right) = \exp\left[- \sum_{n=1}^{N}\int_{x_n}^{x_{n+1}}\rho(\tau) d\tau + \int_{s_I}^{x_{N+1}}\rho(\tau) d\tau \right] \prod_{i=1}^{I} \rho (s_i),
\end{equation}
where $x_n$ refers to the absolute time of the $nth$ laser shot and we let $x_1=0$ for simplicity.  If the data is acquired over many laser shots, we assume that the second integral term in the exponential (a residual accounting for the time between the last observed photon and the end of the image) is negligible.  Using this approximation, we can rewrite equation \eqref{Pois1Dprime} as a 2D image estimation problem where we replace the absolute time variables $s_i$ with time stamps relative to the most recent laser shot $t_i = s_i-x_n$ (for $x_n \le t_i < x_{n+1}$) resulting in
\begin{equation}\label{PDFMulti}
P\left(\left\lbrace \lbrace T_i = t_i \rbrace_{i=1}^{I_n} \right\rbrace_{n=1}^N\right) = \prod_{n=1}^{N} \left\lbrace \exp\left[-\int_{0}^{\Delta t_{RR}}\rho_n(\tau) d\tau\right] \prod_{i=1}^{I_n} \rho_n (t_i)\right \rbrace.
\end{equation}
where $\Delta t_{RR}$ is the repetition period of the laser ($\Delta t_{RR} = x_{n+1}-x_{n}$) and $\rho_n (t)$ is now a 2D variable with continuous dimension in time-of-flight $t$ and discrete dimension in laser shot index $n$.

To employ MLE, we seek to find an estimate $\tilde{\rho}_n(t)$ that minimizes the negative log-likelihood (NLL) of the detection noise model.  This is the optimization loss function that effectively evaluates how well the estimate $\tilde{\rho}_n(t)$ fits the observed time tag data
\begin{equation}\label{TimeTagLoss}
\mathcal{L}_{TT}\left(\tilde{\rho}_n(t); \left\lbrace \lbrace T_i = t_i \rbrace_{i=1}^{I_n} \right\rbrace_{n=1}^N\right) = \sum_{n=1}^N \left ( \int_{0}^{\Delta t_{RR}}\tilde{\rho}_n(\tau) d\tau - \sum_{i=1}^{I_n}\ln\tilde{\rho}_n(t_i)\right).
\end{equation}

Equation \eqref{TimeTagLoss} represents the generalized fit loss for estimating a photon arrival rate $\tilde{\rho}_n(t)$ from observations of time tag data.  This provides a direct mechanism for evaluating the accuracy of candidate solutions for $\tilde{\rho}_n(t)$. Any variety of basis functions may be used to perform this estimate.  For example, polynomial and spline functions would be reasonable options depending on the specific description of the problem of interest. Later in this work, we will employ PTV, where the basis functions are a set of piecewise constant functions.  This is also a natural description of the standard histogram approach where each bin is treated as a constant over the bin width. Piecewise constant functions will additionally enable further simplification of the NLL.  

As is the case in prior works \cite{Marais2016, Marais2022}, $\tilde{\rho}_n(t)$ may also be a forward modeled function of other retrieved variables in order to obtain denoised estimates of the derived data products rather than the photon arrival rate itself. In this work, we focus on recovering the photon arrival rate because it is an essential component of any quantitative estimate (the retrieved quantity that maps directly onto observations), and therefore many of the concepts covered here broadly apply to a number of lidar architectures and applications.  

\subsection*{Example of Fitting Directly to Time Tag Data}
Here we show a simple example how equation \eqref{TimeTagLoss} can be used to fit directly to time tag data.  We consider a case where the photon arrival rate is known to be a Gaussian (such as in the case of a Gaussian laser pulse scattering off a hard target) with background and thus the captured photon flux has the form
\begin{equation}\label{sim1Drho}
\rho(t) = A\exp\left( -\frac{(t-\mu)^2}{2\sigma^2} \right) + b
\end{equation}
Because the form of $\rho(t)$ is known, $\tilde{\rho}(t)$ can be obtained by estimating $A$ (amplitude), $\mu$ (mean), $\sigma$ (standard deviation) and $b$ (background).  Note that in this case, we can take advantage of the fact that the integral of $\rho(t)$ has an analytical definition which is needed in equation \eqref{TimeTagLoss} for MLE estimation of a Poisson point process 
\begin{equation}
\int_0^t\rho(\tau) d\tau = \frac{1}{2} A \sigma \sqrt{2\pi} \left[1 + \mathrm{erf} \left(\frac{t-\mu}{\sigma\sqrt{2}}\right)\right] + b t
\end{equation}
where $\mathrm{erf}$ is the Gauss error function.

\begin{figure}[ht]
\centering\includegraphics[width=12cm]{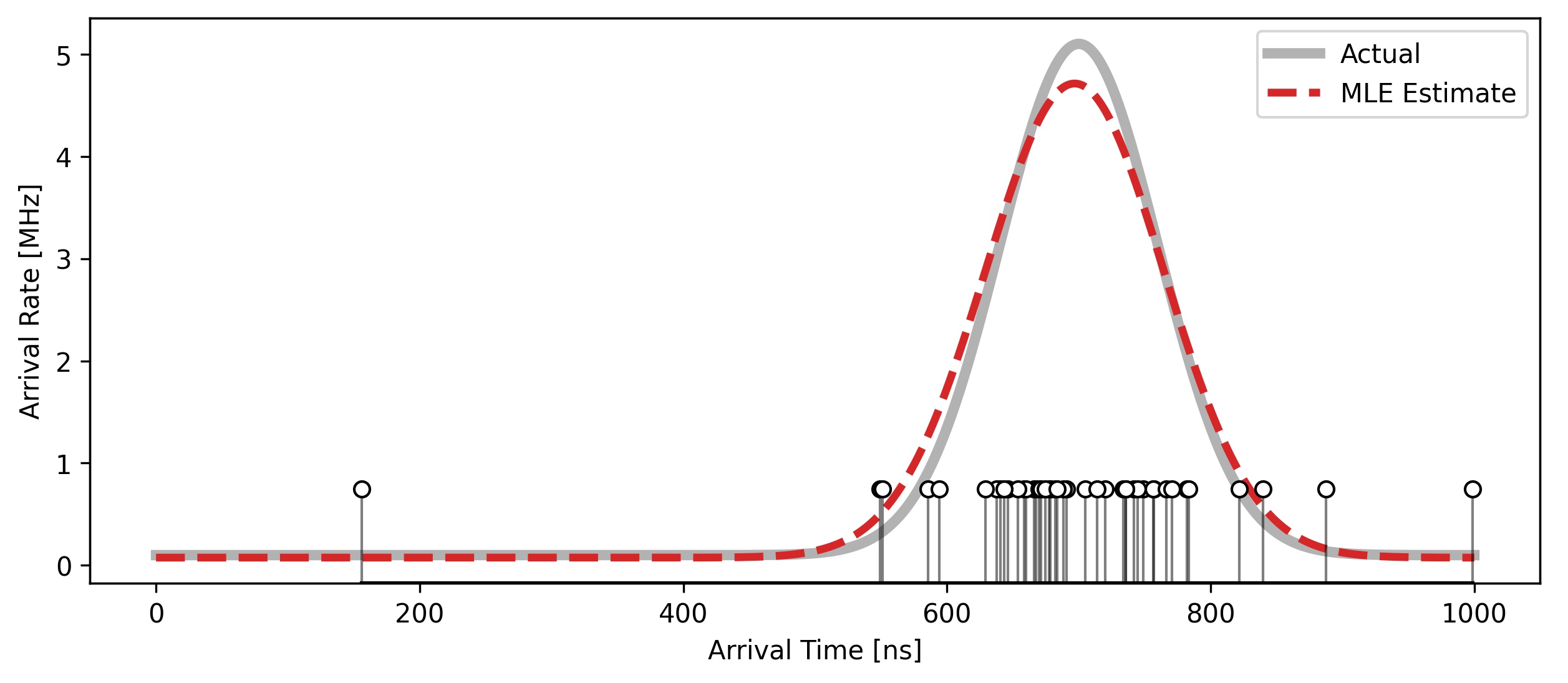}
\caption{Simulation of raw time tags (stems), Gaussian fit to time tags (red dashed) and actual photon arrival rate used to generate the time tags.}
\label{fig:sim1D}
\end{figure}

As a simple demonstration of the concept, we generated simulated time tag data for the non-homogenous point process parameterized by $\rho(t)$ over 50 laser shots.  The actual value of $\rho(t)$ is shown as the gray line in Figure \ref{fig:sim1D} and the simulated photon time tags are shown as the stems.  A photon flux estimate $\tilde{\rho}(t)$ is obtained by minimizing the Gaussian parameterization of $\rho(t)$ in equation \eqref{TimeTagLoss} and the result is shown as the red dashed line.  Note that for greater laser shots, the estimate would become more accurate.  However, the purpose of this example is to demonstrate that fitting can be performed on the time tag data directly.

The theoretical benefit of fitting to time tags is that it eliminates discretization errors when fitting continuous functions to observed time tag data.  This allows one to maximally leverage the theoretical concept of time tag observations.  However, due to the fact that acquisition systems inevitably have resolution limits dictated by the acquisition clock speed, in practice, the distinction between time tag data and high-resolution histograms is not always significant (e.g. in \cite{Rapp2019} data is placed in histograms for processing).  However, there may be cases where data is sufficiently sparse, that fitting directly to time tags is the most practical solution.

\subsection*{Holdout Cross Validation}
In the previous example, we had full knowledge of the functional form of the backscatter signal and could therefore use that knowledge to choose a reasonable basis set of fit parameters. That is not true of most atmospheric measurement problems of practical interest. Therefore, in most cases, we have to choose more general basis sets and trade model complexity and fit quality.  Insufficient complexity will fail to accurately capture the functional form of the backscatter signal, while excessive complexity will tend to erroneously fit noise.  Both instances result in an excessive error in the estimate. Non-quantitative tuning of retrievals, for example, based on expert judgement or past experience (e.g. smoothing according to the researcher's discretion of what ``looks good''), is susceptible to both types of errors while adversely affecting the repeatability of processing tuning.

To address this issue, holdout cross-validation can be used to optimize the signal processing tuning parameters (complexity) to the immediate scene \cite{Hayman2020}.  For example, bandwidth of lowpass filters, polynomial order or total variation regularization can all be optimally determined by evaluating the fit solution against observations with the same mean but statistically independent noise.  This uses the same NLL function as in the fitting process --- where the observations are replaced with independent validation data \cite{Haste}. This allows for an analytical comparative estimate of error and therefore optimization of the basis set complexity for an arbitrary scene with characteristics that are not known \emph{a priori}. 

To methods can be employed for obtaining holdout data from photon counting lidar data.  In the case of Poisson distributed observations, fit and validation data can be split out from the original dataset using Poisson thinning \cite{Oh2013}.  In the case of histograms, Poisson random data is split using a binomial random number generator as described in \cite{Hayman2020}.  Thinning time tag data resulting from a Poisson point process is also possible using a Bernoulli random number generator for each time tag.  Each time tag is placed into a set according to the outcome of that Bernoulli trial (e.g. 0 - fit data, 1 - validation).  We should note that the process of thinning will reduce the amount of signal available for direct processing, but we have found that signal processing that is analytically optimized for the scene seems to always produce better results than the alternative.

In this work, we employ ``manual'' thinning to generate a validation dataset.  In this approach, data from every other laser shot is split between a fit and validation dataset.  This is possible because the laser sample rate (in this case 8 kHz) is much higher than the resolvable variations in the observed scene (in this case on the order of 10s to 100s of Hz).  However, this highlights the potential drawback of manual thinning if the scene is not oversampled by the laser, i.e. from fast-moving platforms, in highly dynamic scenes or lower repetition rate lasers.  However, the benefit of the manual thinning approach is that it allows us to apply holdout cross-validation even when the noise model is not Poisson, which is potentially an important application for the methods described here.  For instances of analog detection or where detector nonlinear effects result in non-Poisson photon counts, manual thinning still presents a viable option for holdout cross-validation.

\section*{Poisson Total Variation}

Poisson Total Variation is a signal processing approach that draws inspiration from medical imaging techniques \cite{Harmany2012}.  The technique was first applied to lidar data for High Spectral Resolution Lidar (HSRL) retrievals of backscatter and extinction coefficients \cite{Marais2016} and later adapted to obtain denoised retrievals of water vapor profiles from MicroPulse DIAL (MPD) data \cite{Marais2022}.  Under the PTV approach, variables are represented as images consisting of discrete pixels, which we assume can be reasonably approximated using a basis of piece-wise constant functions.  This solution basis is imposed by applying total variation regularization to the estimated lidar signal, which is a penalty for each change in value between pixels (i.e. increase in complexity) in the estimated image.  In concept, this forces the optimizer to prefer simpler images, where responding to a change in signal must produce a sufficient gain in fit quality to overcome the penalty of the change. Under this framework, the minimized objective function becomes a sum of the noise model and the total variation regularization
\begin{equation}\label{objective_fun}
\mathcal{O}(\bm{x};\bm{y}) = \mathcal{L}\left[f(\bm{x});\bm{y}\right] + \eta ||\bm{x}||_{TV}
\end{equation}
where $f(\bm{x})$ is the forward model that maps the estimated variable image $\bm{x}$ to a photon flux image $\bm{\tilde{\rho}}$, $\bm{y}$ are the 2D photon count observations, $\eta$ is a scalar tuning parameter that sets the amount of total variation penalty for the 2D image $\bm{x}$ consisting of $I$ rows and $J$ columns, and $||\cdot||_{TV}$ is the $l_1$-based, anisotropic total variation defined \cite{Beck2009}
\begin{align}
||\bm{x}||_{TV} = \sum_{i=1}^{I-1} \sum_{j=1}^{J-1} \left(|x_{i+1,j}-x_{i,j}| + |x_{i,j+1} - x_{i,j}|\right) + \nonumber \\
\sum_{i=1}^{I-1}\left|x_{i+1,J} - x_{i,J} \right| + \sum_{j=1}^{J-1}\left|x_{I,j+1} - x_{I,j} \right|.
\end{align}

In order to minimize the objective function, we employ a version of Sparse Poisson Intensity Reconstruction Algorithm (SPIRAL), SPIRAL-TV \cite{Harmany2012,Oh2013}, in combination with Fast Iterative Shrinkage/Threshold Algorithm (FISTA) (specifically fast gradient projection--FGP--in \cite{Beck2009}).  Each optimization step is calculated
\begin{equation}
    \bm{x}^{k+1} = FGP\left\lbrace \bm{x}^{k} - \frac{1}{\tau_k} \nabla \mathcal{L}\left[f(\bm{x}^{k});\bm{y}\right],\frac{\eta}{\tau_k}\right\rbrace
\end{equation}
where $\tau_k$ is effectively the gradient descent step size and determined using SPIRAL-TV.  The acceptance criteria of each step is similarly determined using SPIRAL-TV.  
  
The specific value of the total variation penalty $\eta$ is determined by minimizing the objective function across a variety of values and selecting the solution that minimizes the validation NLL (evaluating $\mathcal{L}(\bm{\tilde{\rho}};\bm{y}_v,N_v)$ where the $v$ subscript indicates independent validation data).  In this way, the regularization is optimized for the specific scene using holdout cross-validation. All fit and tuning parameters are optimized for each scene by the method in a general way, i.e. no human intervention is required.

When we employ PTV, the estimated variable is approximated as a set of piecewise constant functions and requires an assumed discrete grid for the solution basis.  As a result, the solution is defined for discrete intervals of index $p$ in time-of-flight, and $q$ in laser shots, such that $\tilde{\rho}_{n}(t_i) = \tilde{\rho}_{p,q}$ when $t_p \le t_i < t_{p+1}$ and $q \le n < q+1$.  Given the discrete solution, the time tag loss function in equation \eqref{TimeTagLoss} simplifies for PTV to
\begin{equation}\label{PTVLoss}
\mathcal{L}_{PTV}(\bm{\tilde{\rho}}; \bm{y}, N) = \sum_{q=1}^{Q} \sum_{p=1}^{P} \left[ N_q \tilde{\rho}_{p,q} \Delta t_{p} - y_{p,q} \ln \tilde{\rho}_{p,q} \right].
\end{equation}
where there are $P$ time-of-flight bins in the estimated backscatter, with bin width $\Delta t_p$, $Q$ total laser shot bins, and $y_{p,q}$ are the number of photon counts that fell into the $(p,q)$ grid point as defined by its arrival time $t_p$ and laser shot $n$.

Note that the loss function in equation \eqref{PTVLoss} is effectively the loss function employed for a Poisson random number where $\bm{y}$ is a histogram of photon counts.  Because of the discretization in the PTV solution basis, there is no practical reason to retain the time tag data in its raw form. In applying PTV, raw time tag data is naturally arranged into histograms defined by the resolution of the retrieved image $\bm{\tilde{\rho}}$. However, this PTV histogram resolution can be \emph{much} finer than would be typical if we aimed to estimate $\bm{\tilde{\rho}}$ directly from histogram bin counts. This very fine histogram resolution is thus also flexible in responding to the observed scene.

For general instances of time tag estimation using continuous parametric fitting (such as polynomials or splines), the time tag loss function in equation \eqref{TimeTagLoss} can be employed to avoid loss of information caused by binning the photon counts in a histogram.  However, because PTV requires that we select a solution grid size to define the basis image, under the assumption that $\tilde{\rho}$ is constant within a grid point, there is no practical advantage to processing the raw time tags.  Instead, we will use equation \eqref{PTVLoss} when performing PTV retrievals in the next sections.

We will show, however, that data can be recovered at much higher resolutions and more accurately using PTV than when employing standard histogram methods.  The best retrievals from PTV will be obtained from very sparse histogram data that otherwise has no immediately discernible signal using the standard approach at the same resolution.

\subsection*{Coarse-to-Fine}
Directly processing fine-resolution (sparse) photon count data with PTV does not tend to produce a reliable solution.  In \cite{Marais2016} it is noted that zero-value histogram bins result in non-unique solutions in PTV retrievals.  In previous work, we leveraged a technique termed coarse-to-fine (PTV-CF) in order to obtain more accurate water vapor retrievals over large time intervals \cite{Marais2022}.  Here we show that a similar approach can be leveraged in order to obtain high-resolution estimates of backscatter structure even when the photon counts are small (typically one) and sparse (most bins are zero) at the retrieved resolution.  

\begin{figure}[ht]
\centering
\includegraphics[width=\linewidth]{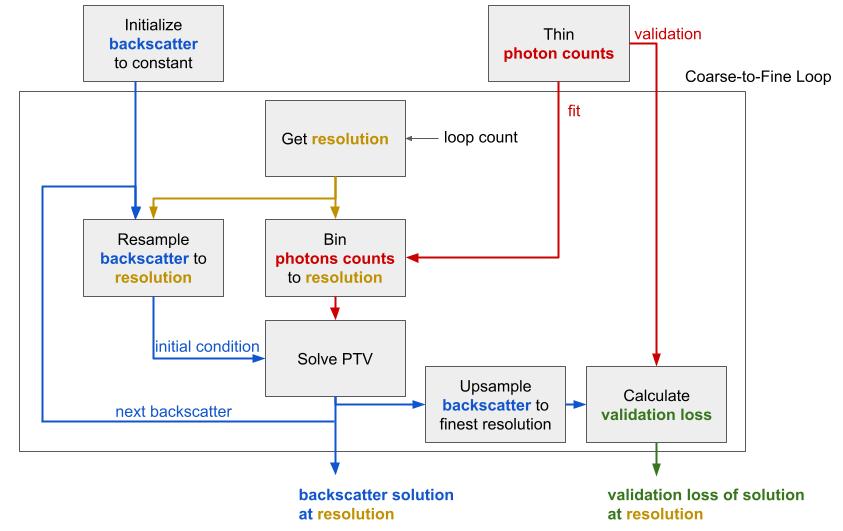}
\caption{Flow diagram of the high-resolution PTV-CF-HR algorithm.}
\label{fig:C2F}
\end{figure}

We term this high-resolution coarse-to-fine version of PTV as PTV-CF-HR.  Simply stated, this processing approach starts with coarse images that have been binned at a low enough resolution to have non-zero photons in all pixels. That first estimate is used as the initial condition for the next estimate at a finer resolution, and the process is repeated.  This approach is described by the flow diagram in Figure \ref{fig:C2F}.  In the figure, key variables are set in color and the arrows define the flow of those variables into each processing step.  Blue corresponds to the backscatter photon flux estimate, yellow is the resolution and the photon counts are red.  We start with a set of photon count time tags and an inverse ordered set of integer multipliers that define the coarse-to-fine resolution steps relative to the base resolution of the final image.  The photon counts are manually thinned by laser shots and binned into a separate fit and validation data set at the finest (final) processing resolution (In the case of the MPD instrument, we allow this to be defined by the rate of the optoelectric switch used to change between transmitting online and offline laser wavelengths).  

The initial estimate of the backscattered photon flux is set to a constant (generally 1 kHz).  For each successive finer resolution, the fit photon counts are binned from the raw time tags to the new finer resolution.  In addition, the initial condition for PTV is the previous solution resampled to the new resolution.  The solution obtained from PTV is evaluated against the validation photon count data by first upsampling the solution to the finest (final) resolution and comparing it (using the NLL) to the validation data similarly binned at the finest base resolution. This final estimate is a refined version of the initial estimate where significant high-resolution features are allowed to persist while noise is largely ignored via total variation regularization.  

In \cite{Marais2022}, coarse-to-fine was performed incrementally only on the water vapor field.  The water vapor estimate was resampled to project onto a base resolution of 1 minute by 37.5 m data.  In this work, we adjust the resolution of the photon count histogram and the retrieved photon arrival rate $\tilde{\rho}$.  In addition, instead of incrementally decreasing bin size (which would require significantly more coarse-to-fine steps), in this work, we typically decrease bin size by a factor of two at each step.

\section*{Simulation}
\begin{figure}[ht]
\centering\includegraphics[width=12cm]{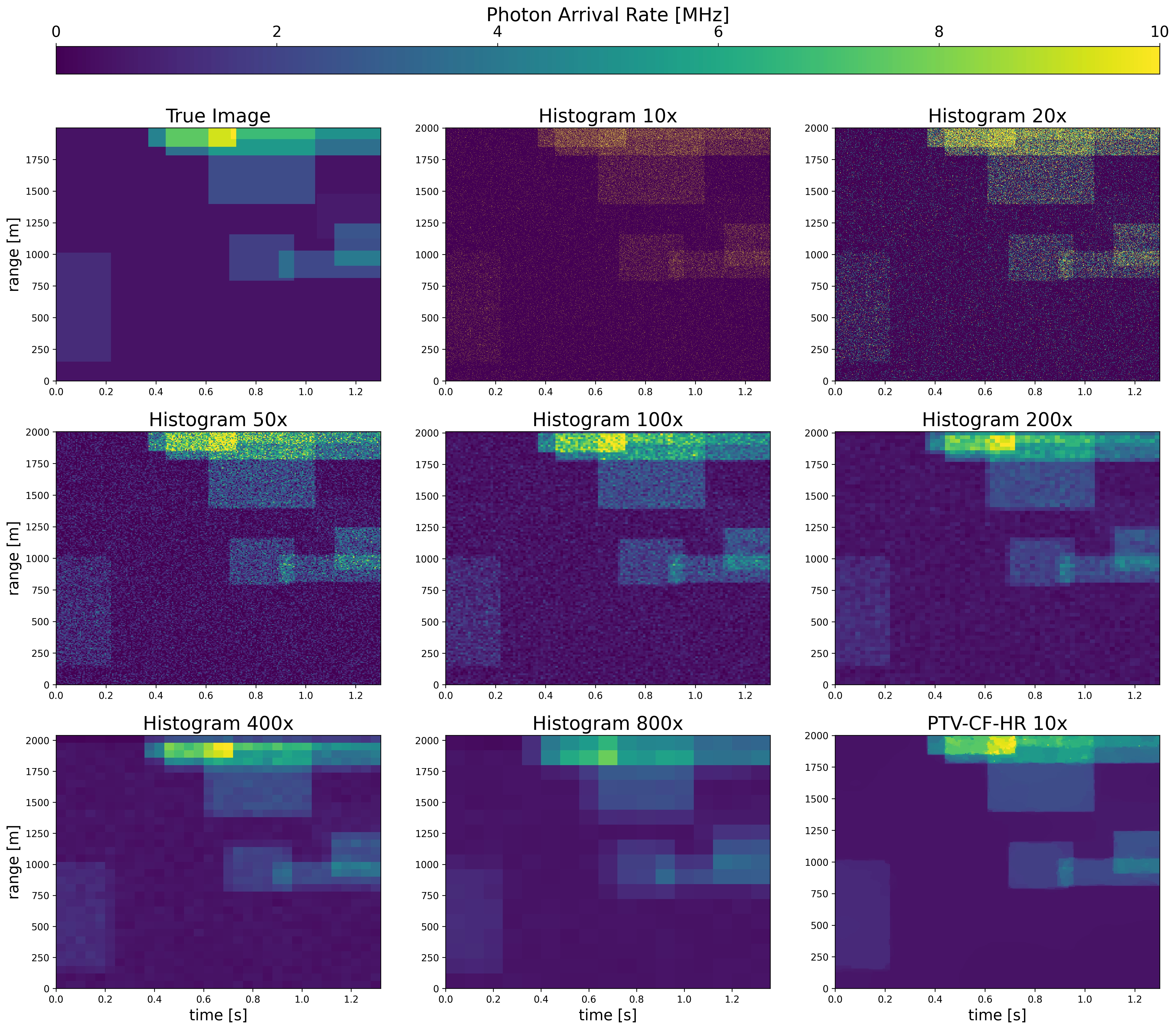}
\caption{True image (top left) compared to histogram estimates of photon flux for incrementally increasing bin size (indicated as multipliers of the base 1 ns range and 0.1 ms time resolution).  At fine resolution, the histograms are noisy while at coarse resolution, images smear out.  The lowest error of the histogram approach is at 200x bin size.  The bottom right image shows the result obtained by PTV-CF-HR at a resolution of 10x.}
\label{fig:sim2D_example}
\end{figure}

In this section, we provide a demonstration of PTV-CF-HR on a simple simulation of randomly generated rectangular patches (where there is overlap, their arrival rates are summed).  This enables us to demonstrate the concept and establish the use of the validation NLL as a metric for evaluating the method on actual atmospheric data in the next section.  This is important because on actual observations no truth is available for computing root-mean-square error (RMSE).  The simulated scene is sampled using a 10 kHz laser repetition rate.  Photon arrivals are simulated as time tagged but then binned to a base resolution of 1 ns (approximately 0.15 m) for PTV processing.  In order to create a validation data set, photon counts from every other laser shot is held out for cross-validation and optimization of the total variation regularizer $\eta$.  An example of the simulated true photon flux is shown in the top left panel of Figure \ref{fig:sim2D_example}.  The fit photon count data is processed at a variety of resolutions which are integer scales of the base resolution (applied to both range and time).  The fit photon count data is placed in a histogram at the same scaled resolution. 

To perform the PTV-CF-HR estimate, we employ the coarse-to-fine solution described in Figure \ref{fig:C2F}.  In this approach, the data is first processed at low resolution using PTV and the solution from that processing step is the initial condition for the next higher-resolution step.  This helps ensure that the solution converges in cases of sparse photon count observations where the optimization problem is not strictly convex.

Figure \ref{fig:sim2D_example} shows the simulated true photon flux along with estimates of the photon flux using a variety of histogram resolutions (reported as multipliers relative to the base resolution) and the result of PTV-CF-HR at 10x the base resolution.  As expected, estimates of photon flux are quite noisy when histogram bins are small.  As we increase the bin sizes, the noise is suppressed and the photon flux estimates improve in the histogram approach.  However, at larger bin sizes, the ability to resolve sharp edges in the image is diminished, resulting in the smearing of the actual structure.  As a result, errors begin to increase beyond a bin resolution of 200x. 
This may be understood as a result of features in the actual image poorly aligning with the predefined histogram grid, or having structure that is finer than the grid.  By contrast, PTV-CF-HR, is largely able to resolve the fine-scale features and edges at 10x resolution, while simultaneously suppressing random noise throughout the image.

\begin{figure}[ht]
\centering\includegraphics[width=10cm]{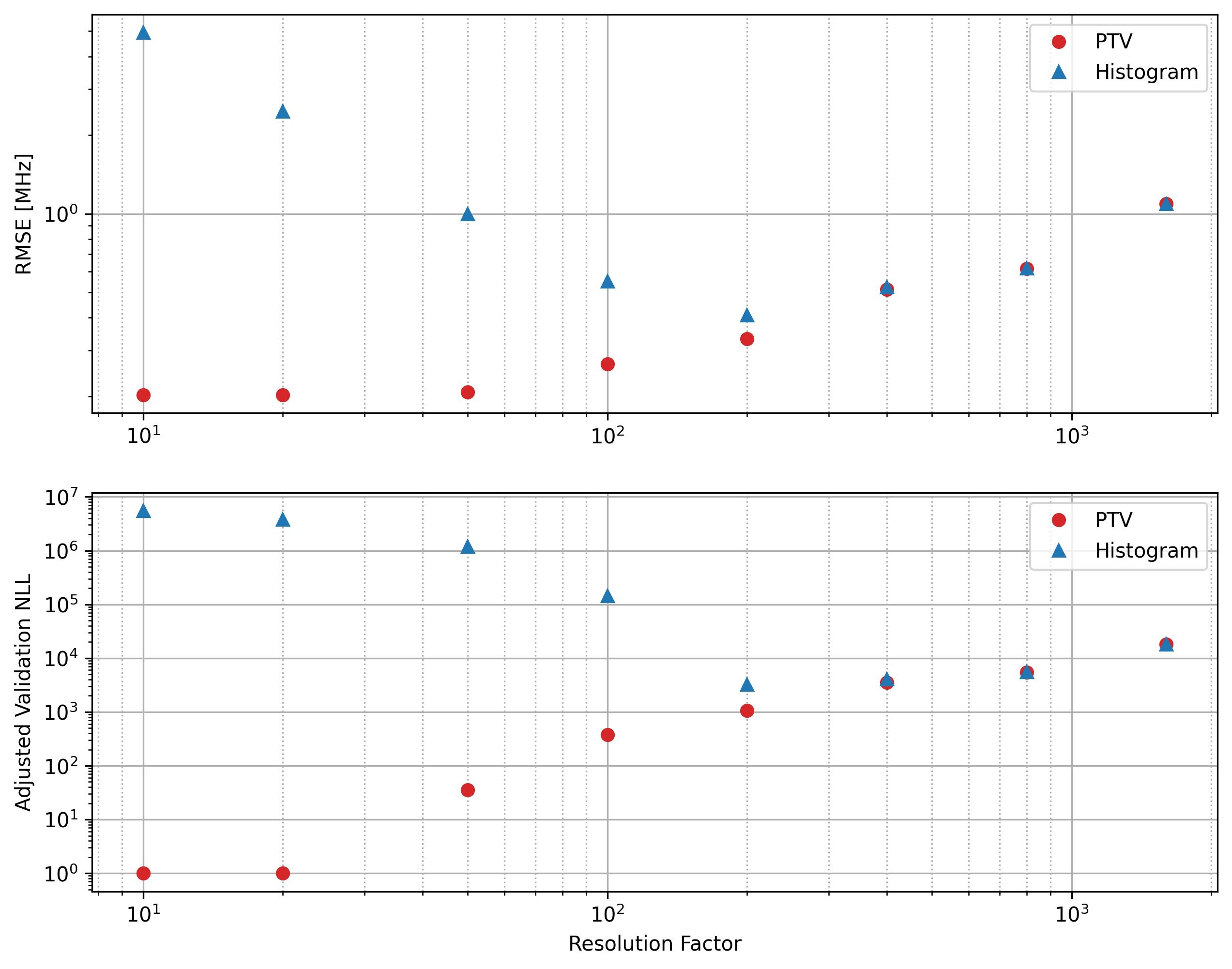}
\caption{Evaluation of PTV processing approach compared to histogram method as a function of data resolution.  RMSE was calculated using the true backscatter flux of the simulation (top). 
 Adjusted validation NLL (bottom) is calculated relative to one plus the minimum NLL.  Resolution is reported as multipliers of the base 1 ns range resolution and 0.1 ms time resolution (larger values indicate coarser resolution).}
\label{fig:sim2D_evaluation}
\end{figure}

We analyze the retrieved estimates of the photon flux using both PTV-CF-HR and the standard histogram approach by calculating the RMSE---which is only possible because we have access to absolute truth in simulation---and the validation negative log-likelihood---which can be applied in actual atmospheric measurement cases as long as we have hold out data.  These error metrics are shown in Figure \ref{fig:sim2D_evaluation} for both estimation approaches at a variety of resolutions (shown as scale factors relative to the base resolution).  In the case of the validation NLL, the actual values are adjusted so the minimum point (which is negative) is at 1, allowing us to display the result on a log axis.  

Note that both the RMSE (top panel) and validation NLL (bottom panel) show similar trends.  This demonstrates how validation NLL can be a useful analytical proxy for error in lidar retrievals where we usually lack a truth reference.  We should note some caveats, however.  The validation NLL is only accurate if the noise model used to calculate the NLL is an accurate representation of the detection statistics, and the noise is uncorrelated between fit and validation observations.  Also, this error is only reflective of how well the retrieval estimate represents the observations.  It cannot account for errors in calibration or inaccuracies in the assumed physical model of the instrument.  

The PTV-CF-HR and histogram approach to estimating photon flux have nearly identical errors at low resolutions (above a scale factor of 400).  For the histogram approach, the lowest RMSE and validation NLL occur at a scale factor of 200 where the two processing approaches have begun to diverge.  At an even finer resolution, PTV-CF-HR continues to become more accurate, due to its improved ability to resolve sharp features in the image, while the histogram approach becomes noisier at finer resolution, and due to this noise, errors begin to increase.  We also note that below a scale factor of 20, there is little additional improvement in the PTV-CF-HR retrieved signal.  This appears to be the performance limit of the retrieval for the given observations.

This analysis shows that by applying MLE techniques like PTV, we can achieve lower error at a higher resolution than the standard histogram processing approach.

\section*{Atmospheric Data}

To further demonstrate PTV-CF-HR, we applied it to atmospheric lidar data and used the validation NLL for evaluation.  The NCAR MicroPulse DIAL (MPD) \cite{Spuler2021} was operated with a custom time-correlated single photon counting (TCSPC) module with 5 ns time-of-flight resolution.  The laser repetition rate is 8 kHz (3 $\mu$J per pulse) with the DIAL system switching between online and offline water vapor channels at a period of 50 Hz.  As a result of this hardware design, to avoid blank laser shots, we chose to bin the data for processing to the finest base resolution of 20 ms (80 shots per wavelength channel) in time and 5 ns (0.75 m) in time-of-flight.

Here we examine data from August 19, 2021 \cite{mpd_data} over a small altitude and time period (approximately 1.9 - 3.3 km for 5 minutes).  During this time we observe falling ice, the melting layer, and subsequent falling rain, which presents an interesting backscatter target for analysis.  All of the photon arrival rates from this scene appear to be well below the nonlinear regime of the single photon avalanche photodiode detector (with a dead time of approximately 25 ns).  

In order to estimate the atmospheric backscatter signal, we employ a forward model that encapsulates the background counts and laser pulse width.  Thus the detected photon arrival rate is
\begin{equation}
\tilde{\rho}_n(t) = L(t) * \hat{\rho}_n(t) + \hat{b}_n
\end{equation}
where $L(t)$ is the square laser pulse (0.625 $\mu$s or 93.75 m) and $\hat{b}_n$ is the background counts estimated using PTV-CF-HR on high altitude photon counts near 18 km (which are assumed to contain negligible atmospheric backscatter signal).  In the results we consider here, the solutions $\hat{\rho}$ are shown in the images, but the validation NLL is still evaluated based on the forward model $\tilde{\rho}$.  In this way, the relatively long square laser pulse can be deconvolved from the atmospheric signal to obtain structure at a much finer resolution.

\begin{figure}[ht]
\centering\includegraphics[width=10cm]{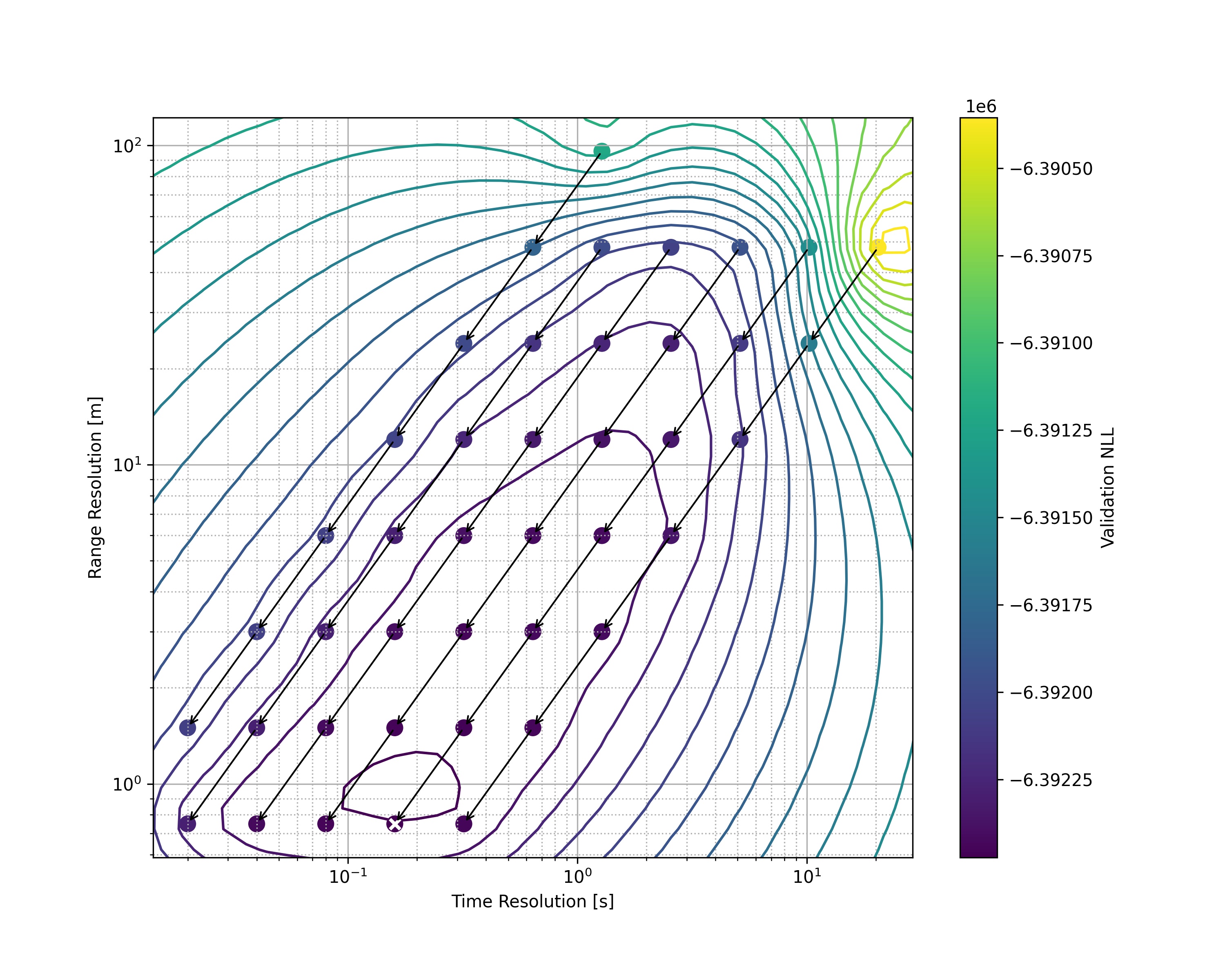}
\caption{Map of validation negative log-likelihood for coarse-to-fine trajectories relating to different time:range bin ratios of the raw photon count data.  The white x indicates the lowest validation NLL of the candidate solutions.}
\label{fig:c2f_trajectories}
\end{figure}

\begin{figure}[ht]
\centering\includegraphics[width=8cm]{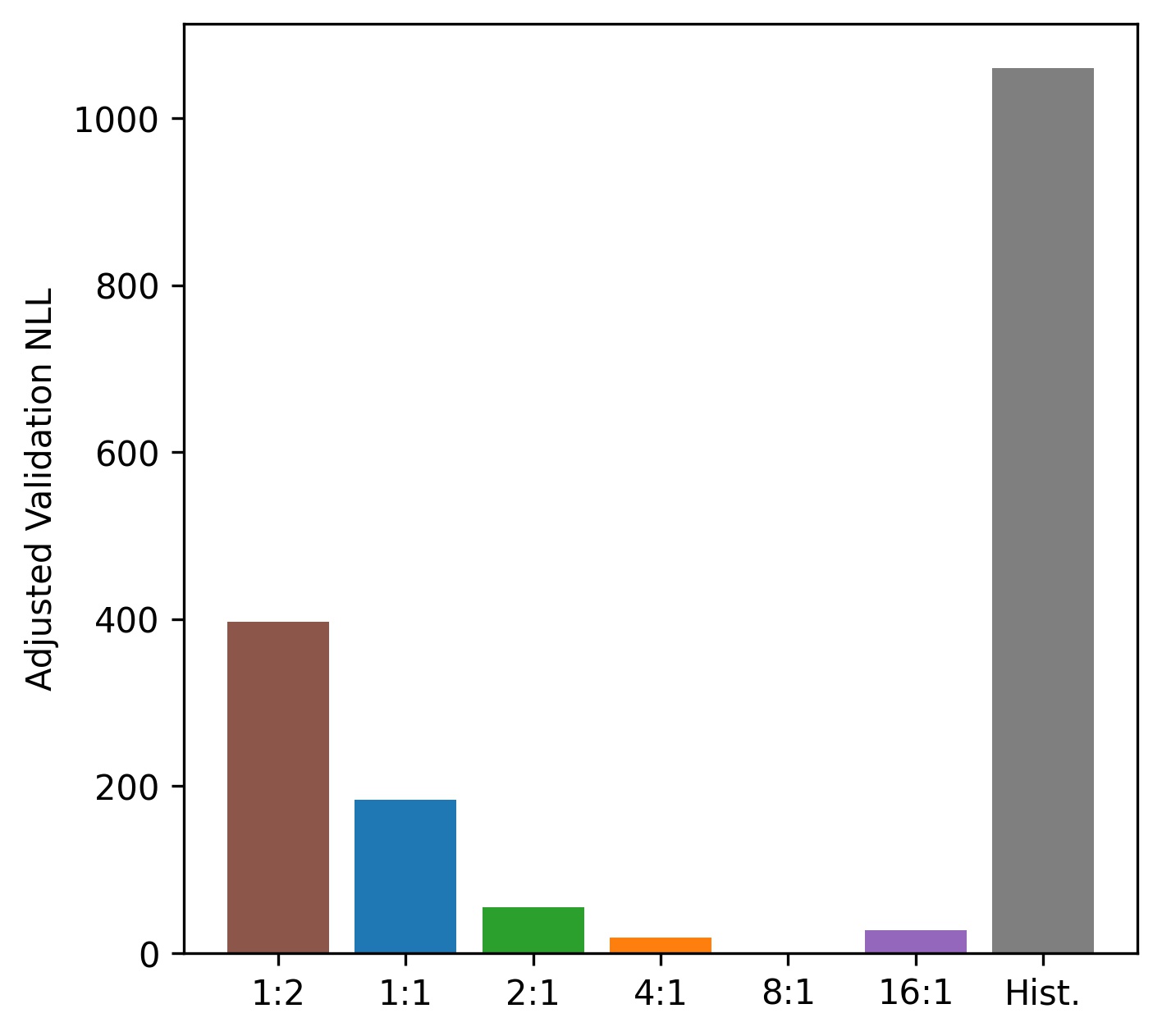}
\caption{Comparison of final estimate validation log-likelihood based on different time:range bin ratios.  Also compared is the optimal histogram binning solution representing the best-case standard approach.}
\label{fig:valdiation_nll_compare}
\end{figure}

In performing this analysis, we consider a few different coarse-to-fine paths, where data is initially binned at different resolution ratios in time and range relative to the base resolution dictated by time of flight resolution (5 ns) and online-offline sampling period (20 ms).  At each up-sampling step, the pixel size was reduced in both dimensions by a factor of two, thus each resolution step is a power of two relative to the base resolution.  The validation NLL of each up-sampled step is shown for each of these time:range processing cases as a function of time and range resolution in Figure \ref{fig:c2f_trajectories}.  The processing result that achieves the lowest validation NLL is 8:1 (bottom row, second from the right, marked with the white x).  At present, we cannot definitively claim to know the reason for this (and therefore have prior knowledge of the best processing option), though it does appear to correspond to the case where the vertical and horizontal pixel dimensions are close to the same.  We compare the final validation NLL of all these processing cases and include the optimal standard approach (Hist.) as shown in Figure \ref{fig:valdiation_nll_compare}.  In order to perform this comparison, we determined the optimal MCS bin resolution based on that which produces the lowest validation NLL.  We note that all of the PTV-CF-HR cases produce a lower validation NLL than the optimal histogram approach.

\begin{figure}[ht]
\centering\includegraphics[width=10cm]{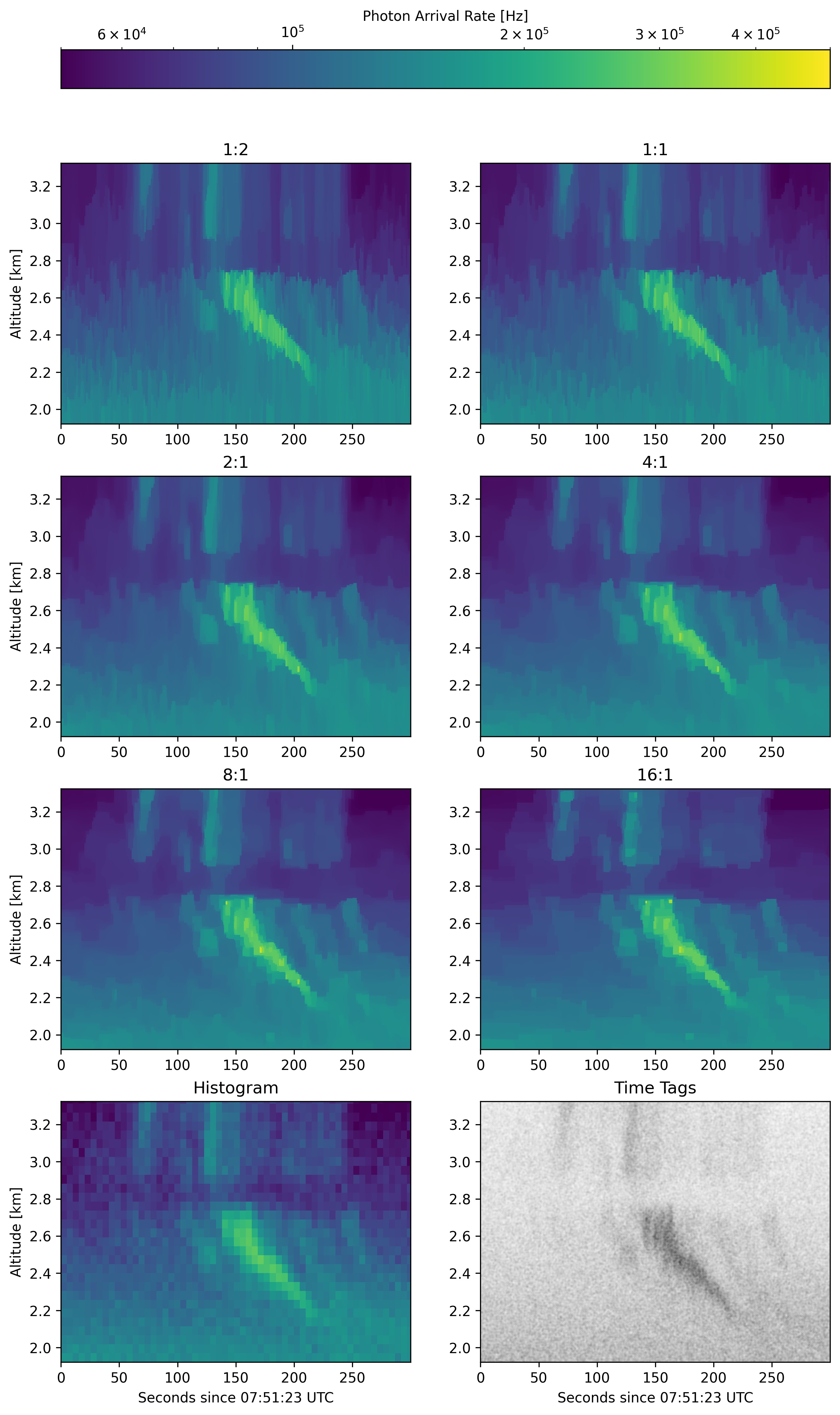}
\caption{Comparison of PTV-CF-HR estimated photon arrival rate based on the relative raw data binning (time:range) compared to the base resolution of 0.02 s x 0.75 m.  Based on validation NLL, the 8:1 case is the best result.  Also shown is the optimal histogram solution representing the best-case standard approach and a scatter plot of the raw photon time tags.}
\label{fig:solution_compare}
\end{figure}

The results of all of these backscatter estimates $\hat{\rho}$ are shown in Figure \ref{fig:solution_compare} which includes the optimal histogram solution (10.24 s x 48 m) and the raw photon time tags as well.  Notably, the optimized histogram solution still retains noise in low signal regimes and significantly degrades the ability to resolve high-resolution structure in higher signal regimes.  We know from the validation NLL in Figure \ref{fig:valdiation_nll_compare}, that this solution has higher error than the PTV-CF-HR solutions, but Figure \ref{fig:solution_compare} makes it visually clear that the standard processing approach results in a lower quality (higher noise, lower resolution) estimate of the lidar backscatter signal.

\begin{figure}[ht]
\centering\includegraphics[width=8cm]{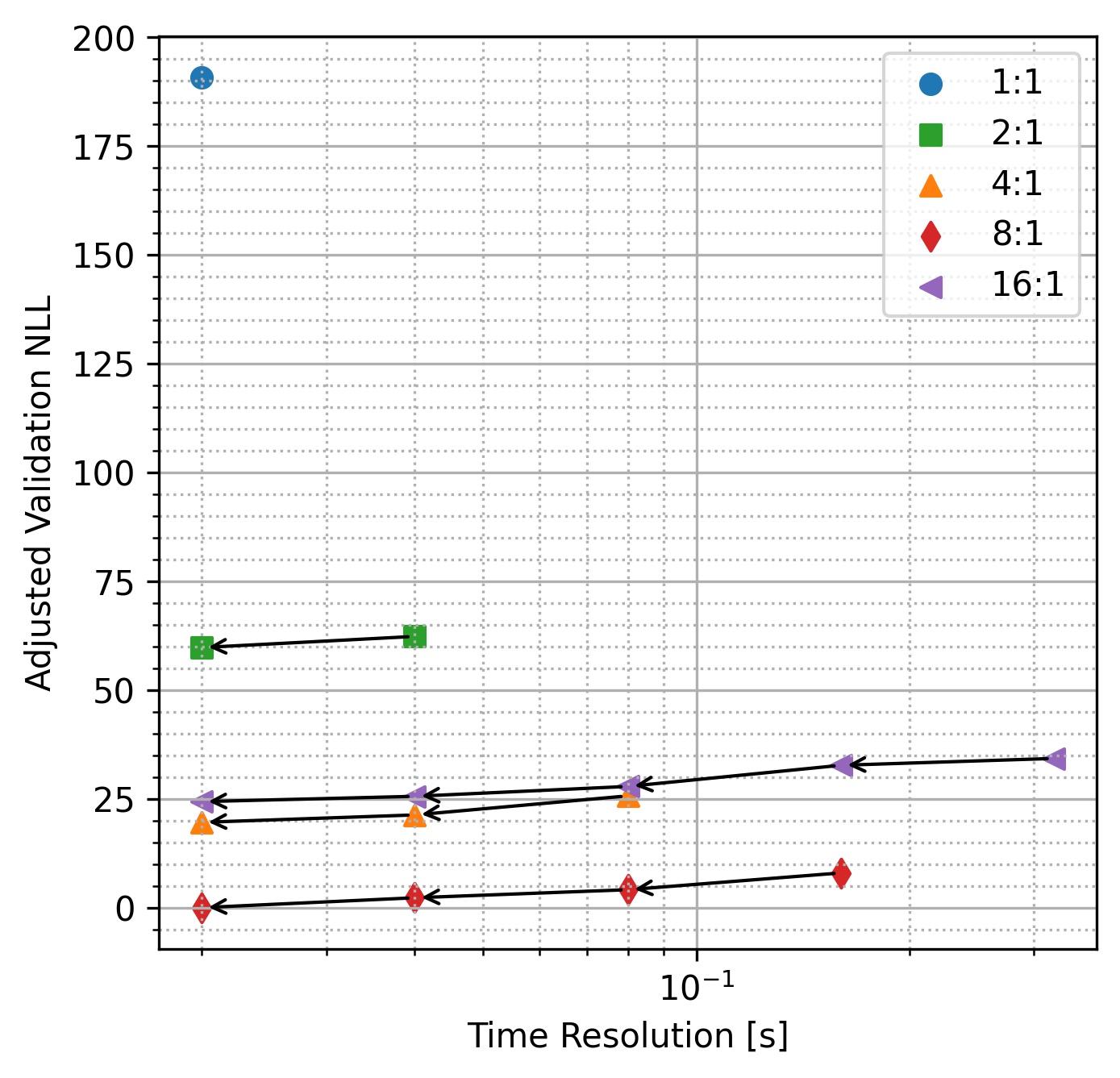}
\caption{Comparison of finest temporal resolution estimates based on initial coarse-to-fine time:range bin ratios.  Processing at finer temporal resolution further improves the validation NLL.}
\label{fig:time_c2f}
\end{figure}

While the 8:1 coarse-to-fine path produces the best final result, there is still potential to process the data at finer temporal resolution. To evaluate the impact of this, we performed coarse-to-fine in only the time dimension.  Figure \ref{fig:time_c2f} shows this analysis applied to all of the cases where we already obtained a minimum in-range resolution (75 cm) as dictated by the acquisition clock.  In this scene, the validation NLL continues to decrease at finer temporal resolution for all of the considered cases.  This suggests that there may be benefits in processing the data at still finer than 20 ms resolution.  However, the impact if this final coarse-to-fine processing is relatively small.  The order of performance remains the same as with the original time:range bin resolutions. 
In most cases, this remains true when comparing the worst and best cases.  For example, note that the 8:1 solution before processing at higher temporal resolutions still has a lower validation NLL than the next best solution (4:1) even at its highest temporal resolution.  This suggests that the initial condition remains an important parameter in obtaining the best possible solution image.  It further emphasizes that, while all of the PTV-CF-HR cases provide superior retrievals to the standard method, more work is needed to determine the optimal initial coarse-to-fine resolution.

It is important to recognize that throughout the analysis of the scene presented here, the validation NLL continues to decrease as the data product resolution becomes finer.  As we move to progressively finer resolutions, we should eventually expect to see the validation NLL cease to improve (or perhaps even become slightly worse) when information captured in the observational data is no longer sufficient to provide higher-resolution observations.  That this does not happen, suggests that the information content captured in this scene supports much higher resolutions than are typically associated with atmospheric lidar data.  That is, in effect, standard lidar resolutions are over-averaging photon counting data.  For the relatively low power MPD, typical time resolutions are 1 minute with a range resolution of 37.5 m.  However, the power of the lidar (and the noise sensitivity of its final data product) will tend to scale the supported resolutions so that higher power systems operating at 1-second resolution, could still be significantly over-averaging as well.  Ultimately, over-averaging is driven by two factors, the spectrum of dynamic/heterogeneous structure in the scene under investigation (potentially extremely high for clouds) and the ability of the captured signals of that structure to rise above noise limits.

\section{Conclusion}
In this study, we have demonstrated a novel approach for applying maximum likelihood estimation to sparse, distributed target data obtained from photon-counting lidar systems.  We have specifically focused on using a coarse-to-fine methodology coupled with Poisson Total Variation, termed PTV-CF-HR, to achieve high-resolution retrievals of atmospheric backscatter.  This was accomplished using low-power lidar instrumentation with raw TCSPC data binned to a minimum temporal resolution of 0.02 s and a range resolution 0.75 m.  

Our findings show that PTV-CF-HR outperformed the standard histogram approach to estimating lidar backscatter signals as shown in both the simulated and atmospheric processing.  Notably, our results reveal that retrievals applied to higher-resolution data have improved validation performance, compared to coarser resolutions.  This result suggests that atmospheric lidar data may generally be over-averaged at resolutions more typical of atmospheric lidar,  although further investigations are needed to understand its impact on data product accuracy.  

While the research presented here has primarily focused on retrievals of backscatter photon arrival rate, it can also be leveraged for atmospheric data products.  This capability has been demonstrated with PTV at much lower resolutions where the data is not sparse \cite{Marais2016,Marais2022}.  Combining the forward model retrieval approach with PTV-CF-HR could be useful for processing low signal-to-noise lidar observations; such as recovering the fine-scale structure of polar mesospheric clouds and sodium layers in the upper atmosphere.  Additionally, it may prove beneficial for retrievals leveraging Quantum Parametric Mode Sorting (QPMS) architectures in environmental sensing where captured signals are likely to be sparse over dynamically evolving scenes.

In addition, PTV-CF-HR serves as a tool for investigating lidar sensor performance and errors resulting from coarse-resolution acquisition and analysis on observational data collected from dynamic scenes.  The effect of averaging lidar signals over heterogeneous cloud structures is rarely discussed in the literature, and there has been very limited investigation of its potential impact.  Some analysis has suggested that space-based lidar products may be biased by heterogeneous scenes\cite{Alkasem2017} and there are similar indications for other remote sensors as well\cite{Arola2022}, but there is currently very limited ability to investigate this effect directly.  Combining PTV-CF-HR with high-resolution acquisition systems (as described in previous works\cite{Barton-Grimley2018,Yang2023} and implemented on the MPD here) offers a pathway to gaining further understanding of remote sensor performance and accuracy.  

While this work shows the potential of Maximum Likelihood Estimation (MLE) in lidar data processing, we also highlight the need for further research.  In particular, we focused on cases where non-ideal detector behavior can be reasonably neglected.  Given that MLE relies on an accurate statistical model of the detection process, achieving accurate recovery of tropospheric cloud structure requires a noise model that can properly account for the non-ideal aspects of the detection processes.  Further, the highest performance result obtained from PTV-CF-HR depended on the initial coarse-to-fine resolution selected.  While it is possible to systematically evaluate a range of possible initial resolutions, this can be computationally expensive.  Additional work that allows us to avoid this step would enhance the practicality of PTV-CF-HR.

\section*{Acknowledgements}
We would like to acknowledge high-performance computing support from Cheyenne and Casper \cite{Cheyenne} provided by NCAR's Computational and Information Systems Laboratory (CISL), sponsored by the National Science Foundation.  This material is based upon work supported by the National Center for Atmospheric Research, which is a major facility sponsored by the National Science Foundation under Cooperative Agreement No.\ 1852977, and NASA 21-NSTGRO22-0326.  We would also like to acknowledge the use of Poisson Total Variation software developed by Willem Marais and his expert advice in developing the solutions described in this work.

\bibliography{main}

\section*{Author contributions statement}
M.H. developed, implemented and analyzed the results of the algorithms and simulations. J.C. Developed the data acquisition hardware for the demonstration. R.S. Integrated the data acquisition hardware with the lidar instrument and acquired the data for processing. G.K., S.S. and J.T. provided technical expertise and advice. M.H. wrote the main manuscript text. All Authors reviewed the manuscript.

\section*{Data availability}
The data used to generate this paper is archived and publicly available \cite{paper_data}.  Atmospheric data is from raw MPD instrument which is publicly available\cite{mpd_data}.  Analysis code used for this publication is available at \url{https://github.com/NCAR/Signal-Estimation-Sparse-Data}.

\section*{Additional information}
The authors have no competing interests to report.

\end{document}